\documentclass[preprint,12pt]{elsarticle}

\usepackage{graphicx}
\usepackage{hyperref}
\input{xy}
\xyoption{all}
\newcommand{\be}{\begin{equation}}
\newcommand{\ee}{\end{equation}}

\journal{Physica A}

\begin{document}

\begin{frontmatter}
\title{Has the world economy reached its globalization limit?}

\author[ift]{Janusz Mi\'s{}kiewicz}
\ead{jamis@ift.uni.wroc.pl}
\author[ulg]{Marcel Ausloos$^{*,}$\footnote{(*) now at 7 rue des Chartreux, B-4122 Plainevaux, Belgium}}
\ead{marcel.ausloos@ulg.ac.be}
\address[ift]{Institute of Theoretical Physics, Wroc\l{}aw University, \\ pl. M.Borna 9, 50-204 Wroc\l{}aw, Poland}
\address[ulg]{GRAPES, ULg., B5a, B-4000 Li$\grave e$ge, Euroland}

\begin{abstract}
The economy globalization measure problem is discussed. Four macroeconomic indices of twenty among the ``richest'' countries are examined.  Four types of ``distances'' are calculated.Two types of networks are next constructed for each distance measure definition. It is shown that the globalization process can be best characterised by an entropy measure, based on entropy Manhattan distance. It is observed that a globalization maximum was reached in the interval 1970-2000. More recently a deglobalization process is observed.
\end{abstract}


\end{frontmatter}

\section{Introduction}
\label{intro}

The term ``globalization'' was used for the first time in Merriam Webster Dictionary in 1944 and is used by economists and in social sciences since the 1960s, e.g. \cite{glob2}. However, this concept did not become popular until the second half of the 1980s. The earliest written theoretical concepts of globalization were penned by an American entrepreneur-turned-minister Charles Taze Russell \cite{glob1}. At present globalization problems are widely discussed: see \cite{scholte2000gci,baylis1997gwp}. There are several aspects of globalization: economic \cite{levitt2005gm,hitt2001smc}, industrial \cite{feenstra1996goa,starr2000nea}, financial \cite{dunning1993gb,maslov-2001-301,mantegna99}, but also political \cite{baylis1997gwp,amin1994gia}, cultural \cite{chossudovsky2005gp,parrenas2001sgw,robertson1992gst}, religious \cite{beyer1994rag} etc... . In fact globalization is expected to result from the growing integration of economies and societies around the world, and is accelerated due to the web and internet.

Within this paper globalization is defined as  {\it the increase of similarities in development (evolution) patterns.} Yet, the key problem is not to prove or disprove here at length the existence of globalization but rather to find whether we can propose an adequate measure of this process, if possible with some universal  aspect. Within this paper four distance measures are thereby tested: two of them are based on straightforward statistical analysis, the other two are based on the entropy concept, through an extension of the Theil index. These four globalization measures are defined in Sec. \ref{sec:glob_mes}. The test data is described in Sec. \ref{sec:data}, i.e. four time series: (i) Gross Domestic Product (GDP), (ii) GDP {\it per capita}, (iii) annual hours worked and (iv) employment {\it per capita}\footnote{The  ``employment rate'' is usually  defined as the ratio between the number of workers to the population size in the 15 - 64 age bracket. However due to the lack of historical data considering the number of people in different age groups the employment ratio is practically taken over the whole country population.},  in the case of ($N= 20$)  ``rich'' countries over more than fifty years. After constructing the distance matrices,  building two types of structurally different networks, considering various time windows and measuring statistical parameters, it occurs that 256 plots should be displayed and discussed.  The results are presented in Sec.\ref{sec:results} through  two subsections based on the time series (i) and (ii) on one hand, and (iii) and (iv) on the other hand. In both subsections, we discuss the results from the point of view of the four distances which are defined in Sec. \ref{sec:glob_mes}. Our own lengthy and detailed examinations suggest us to only display and argue that the most convincing plots pertain to the entropy measure approach coupled to the Manhattan distance study.
Sec. \ref{sec:concl} serves as a section allowing to conclude on statistical and economic findings.
 
\section{Globalization measures}
\label{sec:glob_mes}

Four different distance measures are tested: on one  hand, (i) the correlation distance, Eq.(\ref{eq:corr}), (ii) the mean Manhattan distance, Eq.(\ref{eq:manh}) and on the other hand, entropy based measures, through the Theil index,  i.e.  (iii) the entropy correlation distance,  Eq.(\ref{eq:entr_corr}), and (iv) the entropy Manhattan distance, Eq.(\ref{eq:entr_manh}). The definitions are so given below.
\begin{itemize}
\item (i) Correlation distance 
\be
d_s(A,B)_{(t,T)} = \sqrt{\frac{1}{2} (1- C_{(t,T)} (A,B))}, 
\label{eq:corr}
\ee
based on the linear correlation coefficient $C_{(t,T)} (A,B) $  given by Eq.(\ref{eq:lin_corr})

\be
C_{(t,T)}(A,B) =
  \frac{\langle A B \rangle_{(t,T)} - \langle A \rangle_{(t,T)} \langle B \rangle_{(t,T)} }
{ \sqrt{ \langle A^2 \rangle_{(t,T)} - \langle A \rangle^2_{(t,T)}}
 \sqrt{ \langle B^2 \rangle_{(t,T)} - \langle B \rangle^2_{(t,T)}}
}
\label{eq:lin_corr}
\ee
maps the time series $A(i), B(i)$ onto the interval [0,1], where the discrete index $i$ refers to the time at which some value of e.g. $A$ has been measured. Usually the time series is restricted to a time window ($t,t+T$), where $t$ is the initial point of the time window and $T$ is its width. The brackets $\langle \cdot \rangle_{(t,T)} $ denote the mean value over the interval $(t, t+T)$.

\item (ii) Mean Manhattan distance
\be
\label{eq:manh}
d_l(A,B)_{(t,T)}=\vert \langle  A(i) - B(i) \rangle_{(t,T)} \vert 
\ee
keeping the same notations as in the two previous equations.
\end{itemize} 

Entropy distances are defined in two steps. First the time series are mapped onto an entropy measure, then the distance between such mapped time series are calculated. Thus the procedure implies two possibly different time windows, i.e. an entropy ($T_1$) and a distance ($T_2$) window. Because of the character of the data, limited to a relatively small number of discrete values, the Theil index Eq.(\ref{eq:theil}) is used as the entropy measure \cite{theil1967eai}.

Let us recall the Theil index definition in our context:

\be
Th_A(t,T) =
\label{eq:theil}
 \frac{1}{T} {\sum_{i=t}^{t+T} \left(\frac{A(i)}{\langle A \rangle_{(t,T)}}  \ln \frac{A(i)}{\langle A \rangle_{(t,T)}}\right)},
\ee
 from which the correlation, Eq.(\ref{eq:corr}) or mean Manhattan, Eq.(\ref{eq:lin_corr}) distance can be reformulated, i.e. 
 \begin{itemize}
 \item  (iii) Entropy correlation distance

\be
\label{eq:entr_corr}
d_{se}(A,B)_{(t,T_1,T_2)} =
 \sqrt{\frac{1}{2} (1- C_{(t,T_2)} (Th_A(t,T_1),Th_B(t,T_1)))}
\ee
and 
\item (iv) Entropy Manhattan distance
\be
\label{eq:entr_manh}
d_{le}(A,B)_{(t,T_1,T_2)}=\vert \langle  Th_A(t,T_1) - Th_B(t,T_1) \rangle_{(t,T_2)} \vert .
\ee
\end{itemize}

A matrix of the distances between the various macroeconomic index time series, here in brief called ``countries'', can be next obtained. Notice that if the time window size is shorter than the length of the considered time series then the distance measure can be applied several times, which results in a set of matrix time series for each distance type under consideration. 

Since each distance matrix is an $N \times N$ matrix an appropriate analysis method is required. A network-like structure seems useful. There are several networks to be applied, e.g. Minimum Spanning Tree (MST), which is frequently used in   stock market and other socio-economic topics analyses, e.g. \cite{mst1,mst2,mst3}. It is accepted that MST is very useful in the analysis of dependencies between entities, but in our study we are focused on the general properties of the considered set of countries. Therefore instead of MST networks with clear local rule of attachment are to be prefered. Within this paper, the Bidirectional Minimal Length Path (BMLP) and the Locally Minimal Spanning Tree (LMST) network structures are considered. In short, 
\begin{description}
 \item[BMLP:]
The network begins with, as seed,  the pair of countries with the smallest distance between them. Then the country closest to the nodes being the ends of the (seed) network are searched for and that with the shortest distance is attached to the appropriate end. The algorithm is continued until all countries become nodes of the (linear) network.
\item[LMST:] The root of the network is the pair of closest neighbouring countries. Then the country closest to any node is searched for and attached at the appropriate node, a.s.o., to form a ''tree''.
 \end{description}

The examined data  is next given in Sec. \ref{sec:data}. The statistical parameters of such constructed networks, i.e. mean  (mean) value and standard deviation (std) of the distances between nodes, are calculated and their evolution discussed in Sec. \ref{sec:results}. 

\section{Data}
\label{sec:data}

\begin{figure}
\centering
\includegraphics[bb=50 50 301 226,scale=0.75]{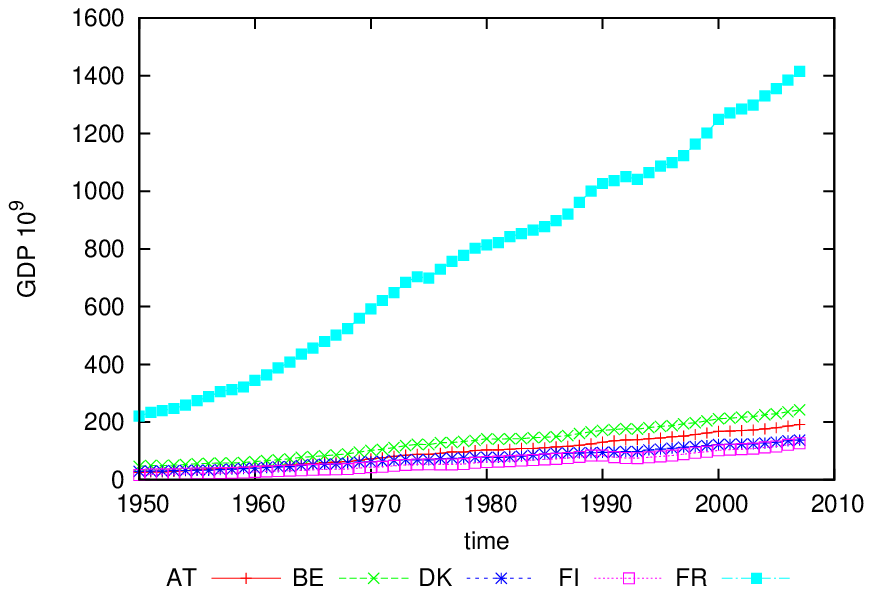}
\includegraphics[bb=50 50 301 226,scale=0.75]{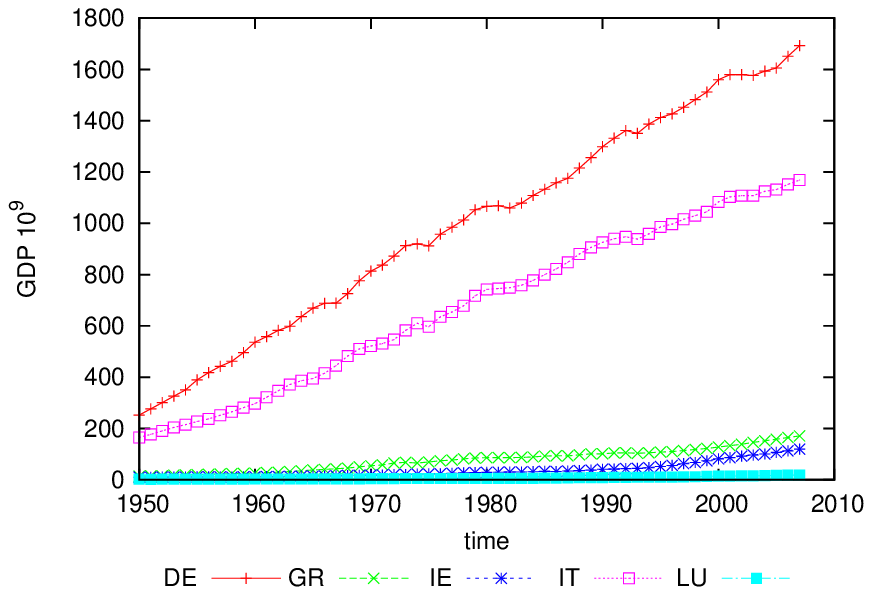}
\includegraphics[bb=50 50 301 226,scale=0.75]{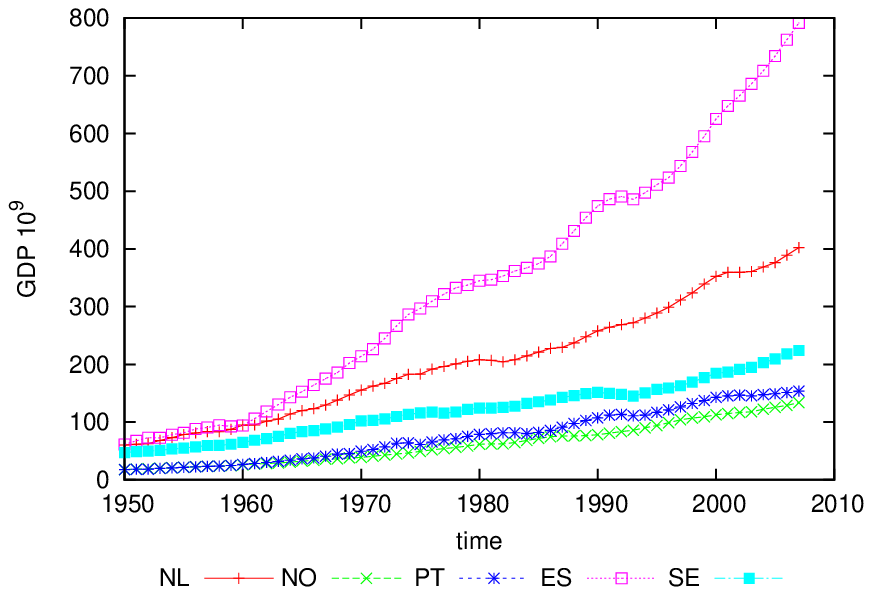}
\includegraphics[bb=50 50 301 226,scale=0.75]{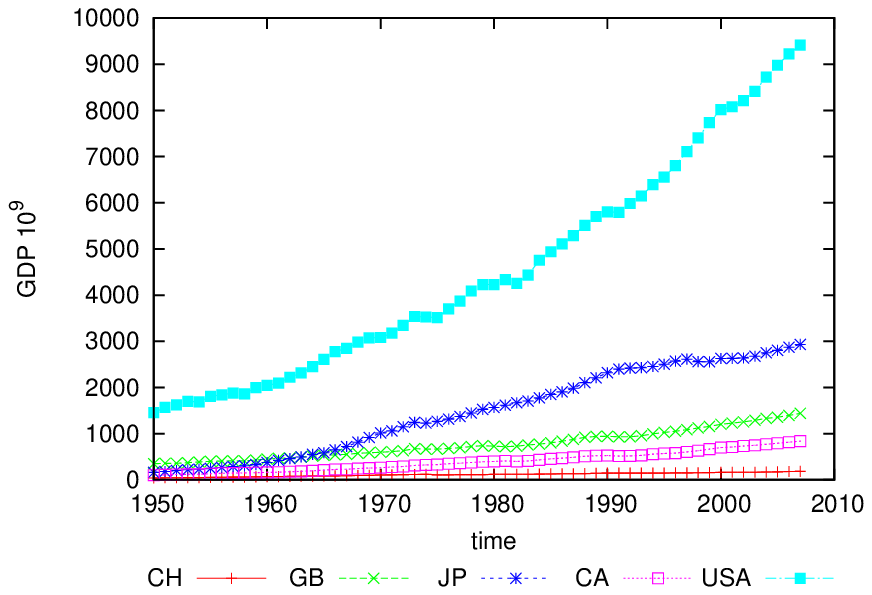}
 \caption{Semilog plots of GDP (in 1990 US \$ units) of the mentioned countries as a function of time }
 \label{fig:gdp}
\end{figure}

\begin{figure}
 %\centering
\includegraphics[bb=50 50 301 226,scale=0.75]{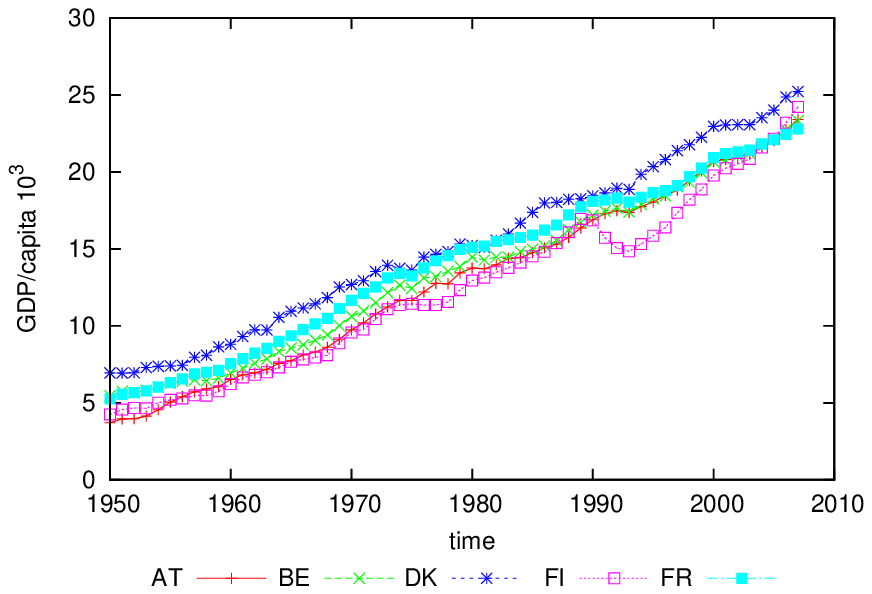}
\includegraphics[bb=50 50 301 226,scale=0.75]{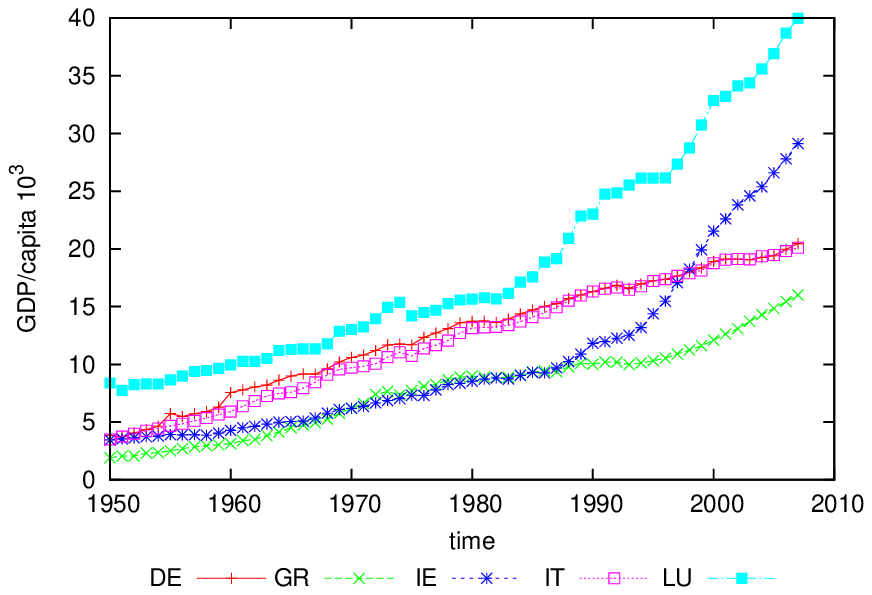}
\includegraphics[bb=50 50 301 226,scale=0.75]{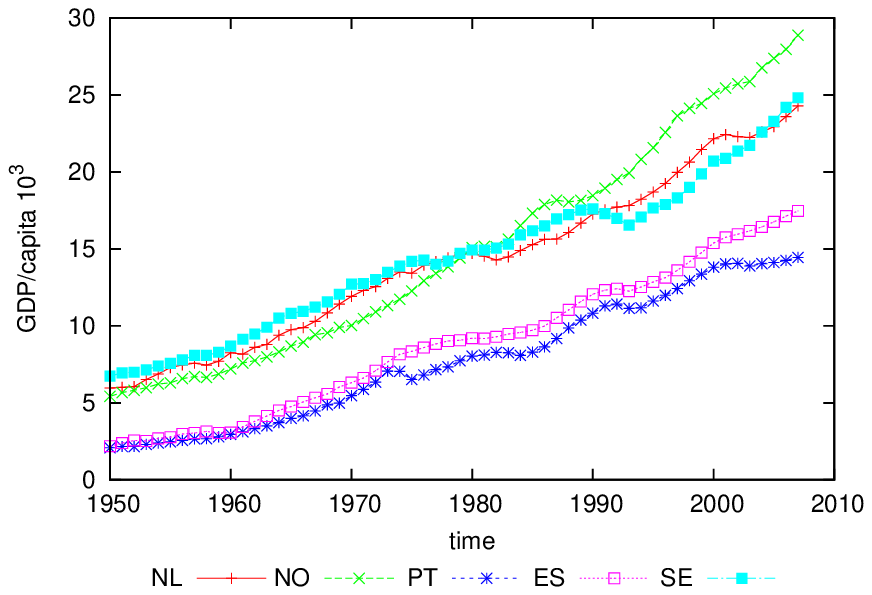}
\includegraphics[bb=50 50 301 226,scale=0.75]{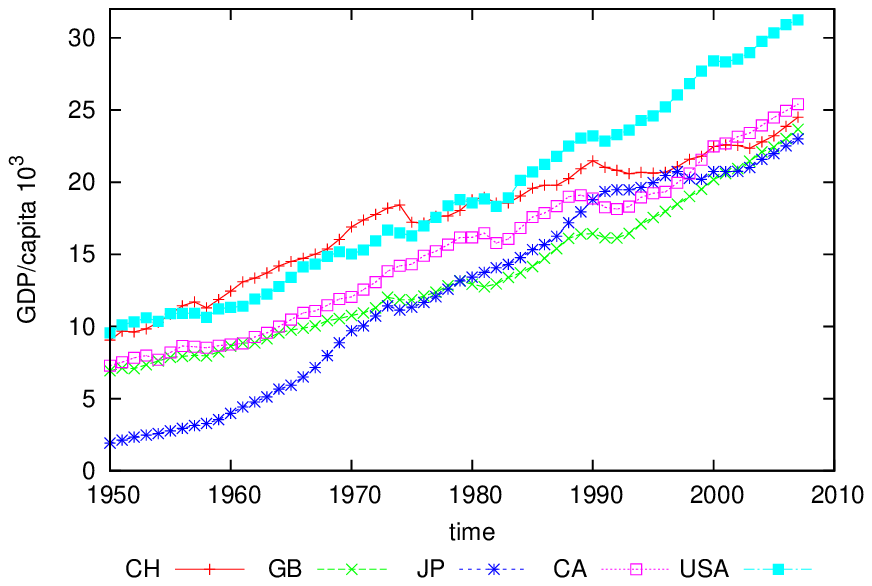}
 % gdp_capita_1.eps: 0x0 pixel, 300dpi, 0.00x0.00 cm, bb=50 50 301 226
 \caption{Semilog plots of GDP {\it per capita} (in 1990 US \$ units) of the mentioned countries  as a function of time}
 \label{fig:gdp_capita}
\end{figure}

\begin{figure}
 \centering
 \includegraphics[bb=50 50 266 201, scale=0.8]{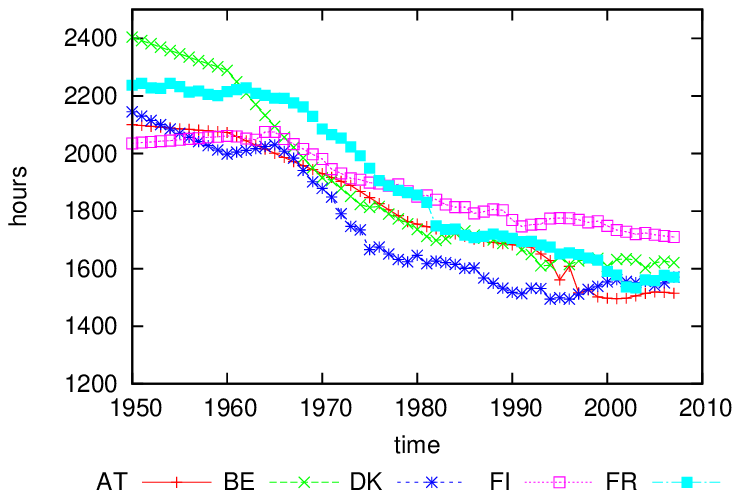}
 \includegraphics[bb=50 50 266 201, scale=0.8]{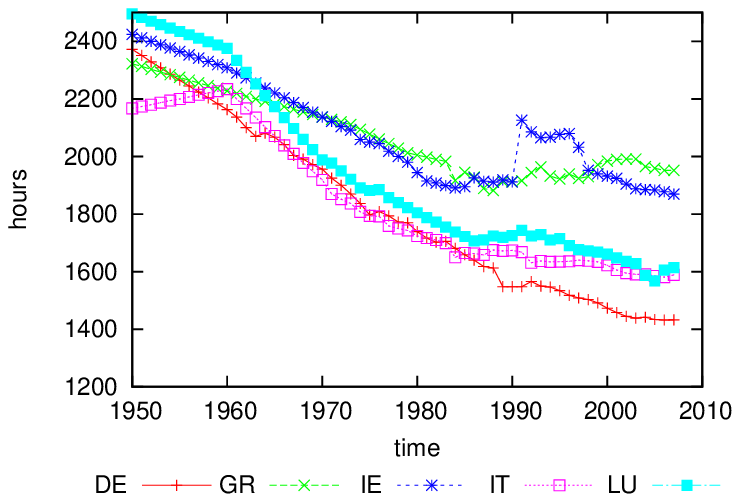}
 \includegraphics[bb=50 50 266 201, scale=0.8]{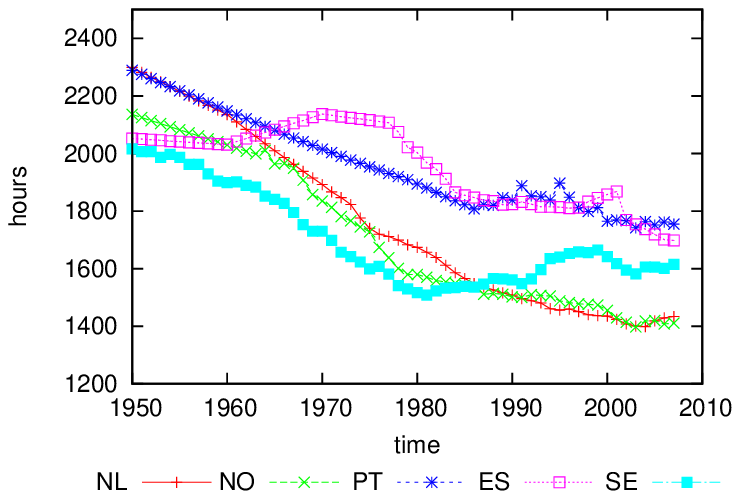}
 \includegraphics[bb=50 50 266 201, scale=0.8]{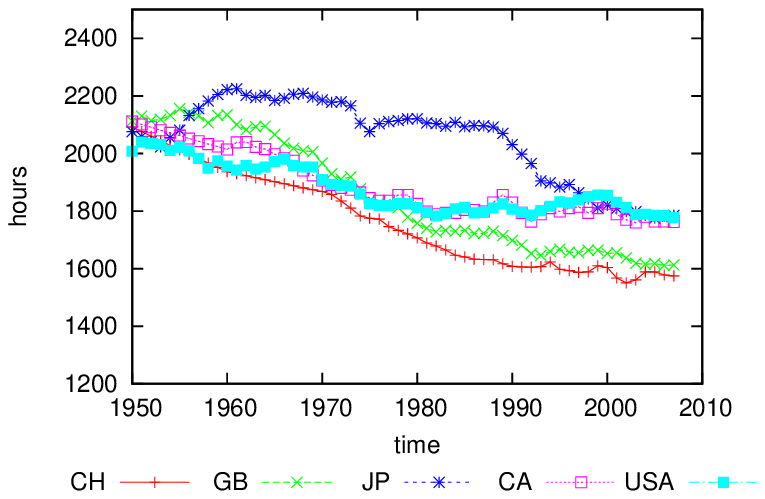}
 % hours_1.eps: 0x0 pixel, 300dpi, 0.00x0.00 cm, bb=50 50 266 201
 \caption{ Annually worked hours of the considered set of countries as a function of time}
 \label{fig:hours}
\end{figure}

\begin{figure}
 \centering
\includegraphics[bb=50 50 266 201, scale=0.8]{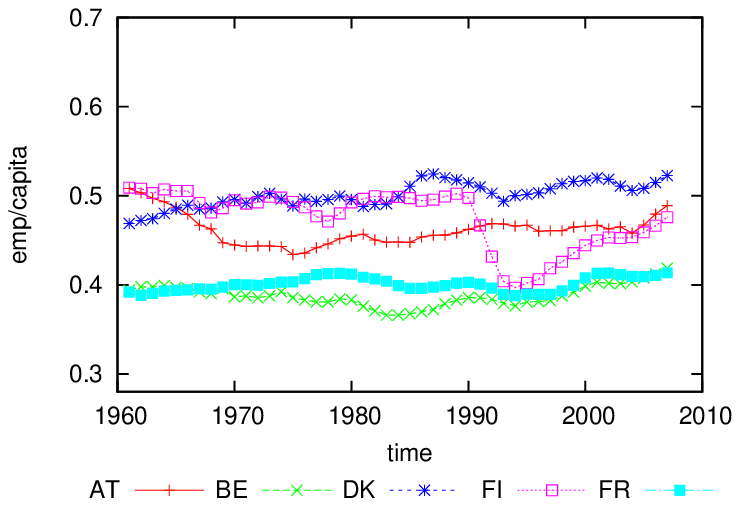}
\includegraphics[bb=50 50 266 201, scale=0.8]{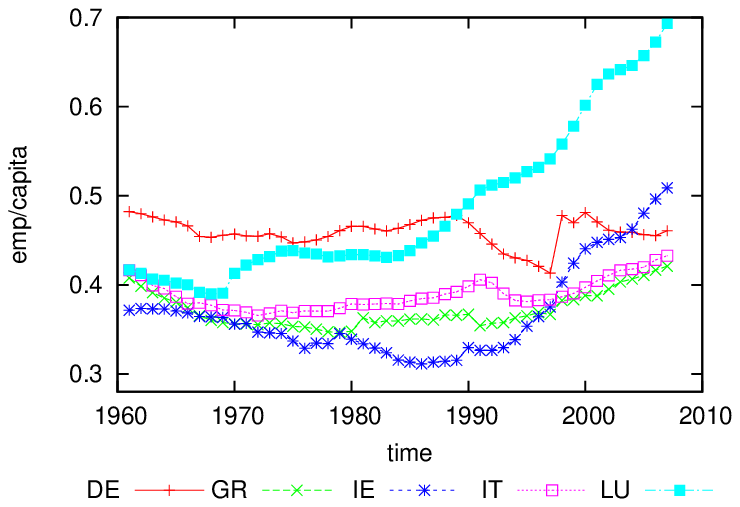}
\includegraphics[bb=50 50 266 201, scale=0.8]{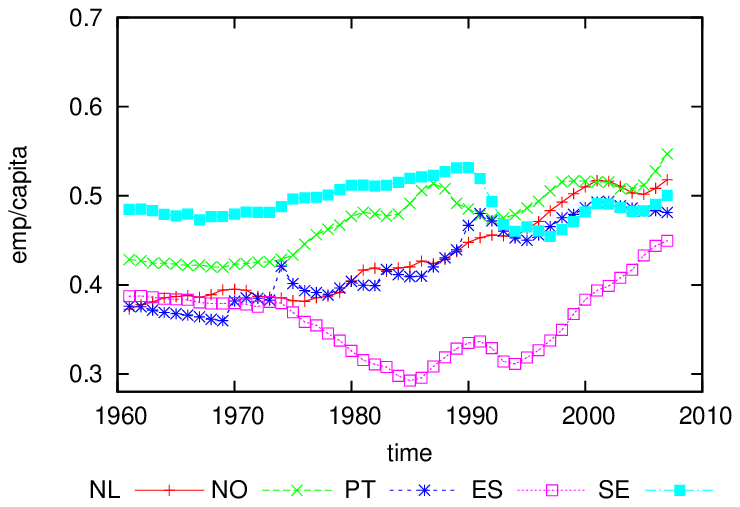}
\includegraphics[bb=50 50 266 201, scale=0.8]{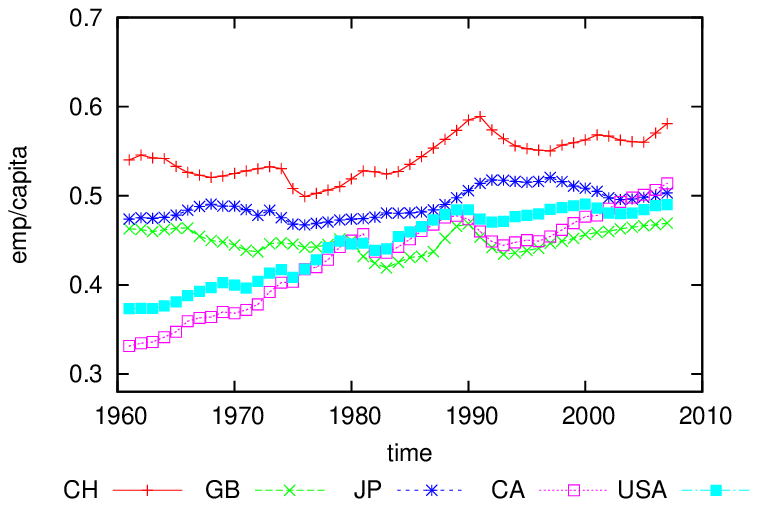}
 % emp_pop_1.eps: 0x0 pixel, 300dpi, 0.00x0.00 cm, bb=50 50 266 201
 \caption{ Employment {\it per capita} ratio of the considered set of countries as a function of time}
 \label{fig:emp_pop}
\end{figure}

The present study is based on the (i) total GDP, (ii) GDP {\it per capita}, (iii) annually hours worked and (iv) employment {\it per capita} ratio as collected by the Conference Board and Groningen Growth and Development Centre \cite{growth2007dca}. The total GDP and GDP {\it per capita} data are given in 1990 US \$ converted at Geary Khamis PPPs and denote the value for a given year \cite{ppp}. Twenty of the most developed countries are analysed: Austria (AT), Belgium (BE), Denmark (DK), Finland (FI), France (FR), Germany (DE), Greece (GR), Ireland (IE), Italy (IT), Luxembourg (LU), the Netherlands (NL), Norway (NO), Portugal (PT), Spain (ES), Sweden (SE), Switzerland (CH), U.K. (GB), Japan (JP), Canada (CA), U.S.A.  (US). The time interval considered spans 57 years, i.e. from 1950 to 2007. 

The GDP data are presented in semilog plots in Fig.\ref{fig:gdp}. The plots indicate a rather steady growth. Moreover one can point out striking similarities between the various GDP evolutions. The similarities are even more self-evident for the GDP {\it per capita} (Fig. \ref{fig:gdp_capita}).

On the other hand the employment market described by annually worked hours (Fig. \ref{fig:hours}) and employment {\it per capita} ratio (Fig. \ref{fig:emp_pop}) are lacking obvious similarities. 
The former has some steady decreasing trend but a general overall rule is hardly found for the latter.

\section{Results}
\label{sec:results}

Globalization analysis are here below reported for the so defined distance measures in the case of four different time window values.
The time windows sizes were chosen such that the averaging procedure should be able to smoothen the noise in order to be able to observe the time evolution of the considered parameters, -- whence the time windows should be as short as possible. In entropy related distances two different time windows are involved; therefore two opposite combinations of the time window size were chosen ($T_1,T_2$).  In so doing  we  can check the role of the time window sizes as well, i.e.
\begin{itemize}
\item for correlation Eq.(\ref{eq:corr}) and for mean Manhattan Eq.(\ref{eq:manh}) distances\footnote{ Within this analysis the time window size is measured in years and simplifying the notations we always write the windows for entropy correlation and entropy Manhattan distances as $(T_1=5 yrs , T_2=10 yrs )$ as $(T_1, T_2)$. }: $T \in \{10,$ $ 15,$ $ 20, $ $ 25 \} $
\item for entropy correlation Eq.(\ref{eq:entr_corr})  and for entropy Manhattan Eq.(\ref{eq:entr_manh}) distances: $ (T_1,T_2) \in \{ (5, 10),$ $ (10, 5), $ $ (10, 10), $ $ (15, 15)  \} $.
\end{itemize}

After the time series were mapped by the appropriate distance measure, the BMLP and LMST networks were constructed and their statistical features analysed. 

Since four different distance matrices are investigated through constructing two network structures, which is done in the case of four types of time series, four time windows, and two statistical parameters are considered, this leads to 256  plots. However for the sake of clarity only the crucial cases are discussed in the following. All   plots are available on request from the authors. For the reader ease we have presented the discussions here below as if based on a logical tree structure, i.e. a first subsection based on the GDP and GDP{\it per capita} time series, followed by a next one with the annually worked hours and employment {\it per capita} ratio. In each subsection  the  discussion follows the order of definition of the distances.

\subsection{Results/discussion: total GDP and GDP {\it per capita} time series}

(i) {\it The correlation distance}:  First consider the case of the {\it total} GDP time series. The extremum points of the mean correlation distance evolution are listed in Table \ref{tab:tot_gdp}. Below we discuss shortly the main features of the mean distance evolution. The mean distance between nodes for BMLP and LMST networks are denoted by $E(d_s \mbox{BMLP})$ and $E(d_s \mbox{LMST}) $ respectively. For the time window $T=10$ the mean distance between nodes for both networks is decreasing from the value $E(d_s \mbox{BMLP})$ $ \approx 0.3,  $ $E(d_s \mbox{LMST}) $ $\approx 0.2$ in 1950\footnote{The date corresponds to the initial point of the time window, i.e. in the case of the time window $T=10 $ the time window which begins on the 01 of January 1950 ends on the 31 of December 1959. For simplifying the notations we always write the windows as $(T_1, T_2)$. } to the level $E(d_s \mbox{BMLP}) \approx E(d_s \mbox{LMST}) \approx 0.05$ in 1957. The mean value remained on this level until 1964 and increased up to $E(d_s \mbox{BMLP}) \approx E(d_s \mbox{LMST}) \approx 0.45$ in 1970, then was rapidly decreasing to achieve a local minimum in 1975 at the mean value  $E(d_s \mbox{BMLP}) \approx 0.15, $ $ E(d_s \mbox{LMST}) \approx 0.12$. After a relatively stable evolution in [1975-1980], a second maximum is observed in 1986 with  $E(d_s \mbox{BMLP})$ $ \approx 0.5, $ but  $E(d_s \mbox{LMST}) \approx 0.35$. This maximum is followed by a significant decrease down to $E(d_s \mbox{BMLP}) \approx E(d_s \mbox{LMST}) \approx 0.15$ in 1990. For the time window $T=15$ the mean value of the distances between nodes is slowly decreasing from $E(d_s \mbox{BMLP})$ $ \approx 0.11$,  and $E(d_s \mbox{LMST})$ $ \approx 0.10$ in 1950 till $E(d_s \mbox{BMLP})$ $ \approx 0.06, $ and  $ E(d_s \mbox{LMST}) \approx 0.05$ in 1959.

 From that time on, the mean distance is growing till 1970 $[E(d_s \mbox{BMLP})$ $\approx  $ $ E(d_s \mbox{LMST})$ $ \approx 0.25]$. In the case of the BMLP network the next minimum is observed in 1975 $E(d_s \mbox{BMLP})$ $ \approx 0.13$ before it  increases toward a stable value in [1980-1986], i.e.  $E(d_s \mbox{BMLP}) \approx 0.25$.  In the last interval of the considered period the mean distance of the BMLP network is decreasing. For the LMST tree the mean distance between countries remains in the interval $0.1 < E(d_s \mbox{LMST}) <0.15$ after the local minimum in 1975.

For the time window $T=20$ for both network structures the mean distance between countries (or nodes) remain in the interval $0.05<E(d_s \mbox{BMLP})$ $ < 0.22$, and $0< E(d_s \mbox{LMST}) < 0.15$.

For the longest, considered here time window $T=25$ the evolution of the mean distance between nodes is similar to the one  observed for the case $T=20$ and takes values in the intervals $0.07<E(d_s \mbox{BMLP}) < 0.20$, and $ 0.05< E(d_s \mbox{LMST}) < 0.12$ respectively.

Next consider the second set of time series, i.e. the GDP {\it per capita}. The extremum points of the mean correlation distance  evolution in the case of LMST network are listed in Table \ref{tab:gdp_per_capita}.  The mean distance between nodes for BMLP network, except for the time window $T= 10 $, increases almost monotonically in the considered interval. For the LMST network and $ T=10, $ the mean distances decrease in the interval 1965 till 1981 from $E(d_s LMST)=0.7$ to $E(d_s \mbox{LMST})=0.41$ and increase to a local maximum in 1993 $E(d_s \mbox{LMST})=0.57$. In the case of $T=15, $ a decrease of $E(d_s \mbox{LMST})$ period can be pointed [1970, 1985], which is followed by an increase of $E(d_s \mbox{LMST})$ till 1990. The local maxima and minima are: $E(d_s \mbox{LMST})(1970) = 0.61$, $E(d_s \mbox{LMST})(1985) = 0.45$, $E(d_s \mbox{LMST})(1990) = 0.60$. 

For the time window $T=20, $ the first local maximum is observed at 1960  $E(d_s \mbox{LMST})(1960) = 0.64$; then the distance is decreasing to $E(d_s \mbox{LMST})(1967)$ $ = 0.48$ and  next increasing to  $E(d_s \mbox{LMST})(1970) = 0.58$  before  finally decreasing to $E(d_s \mbox{LMST})(1987) = 0.47$.  

For the time window $T=25, $ the initial evolution of the mean distance is scattered around $E(d_s \mbox{LMST}) = 0.56$; thereafter, the mean distance is decreasing from 
$E(d_s \mbox{LMST})(1970) = 0.57$ to $E(d_s \mbox{LMST})(1987) = 0.46$.

\begin{table}
\begin{center}
% use packages: array
\begin{tabular}{|l|l|l|c|c|c|c|c|c|}
\hline
&window &size &  \multicolumn{3}{|c|}{10 } & \multicolumn{3}{c|}{15 }  \\ \hline
B & local & value & 0.3 & 0.45 & 0.5 & 0.11 & 0.25 & 0.25 \\ \cline{3-9}
M & max & year & 1950 & 1970 & 1986 & 1950 & 1970 & 1980 \\ \cline{2-9}
L & local & value & 0.05 & 0.15 & 0.15 & 0.06 & 0.13 &-  \\ \cline{3-9}
P & min & year & 1957 & 1975 & 1990 & 1959 & 1975 &-  \\ \hline
L & local & value & 0.2 & 0.45 & 0.35 & 0.1 & 0.25 &  -\\ \cline{3-9}
M & max & year & 1950 & 1970 & 1986 & 1950 & 1970 &-  \\ \cline{2-9}
S & local & value & 0.05 & 0.12 & 0.15 & 0.05 &  -&  -\\ \cline{3-9}
T & min & year & 1957 & 1975 & 1990 & 1959 &  -&-  \\ \hline
\end{tabular}
\end{center}
\caption{The local extrema of the mean distance between the 20 rich countries total GDP,  in the case of  the correlation distance   definition and on BMLP and LMST networks.
\label{tab:tot_gdp}}
\end{table}

\begin{table}
\begin{center}
% use packages: array
\begin{tabular}{|l|l|c|c|c|c|c|c|c|}
\hline
window & size &  \multicolumn{2}{|c|}{10 } & \multicolumn{2}{c|}{15 } & \multicolumn{2}{c|}{15 } & 25   \\ \hline
local & value & 0.7  & 0.57  & 0.61  & 0.6  & 0.64  & 0.58 & 0.57 \\ \cline{2-9}
max & year & 1965 & 1999 & 1970 & 1990 & 1960 & 1970 & 1970 \\ \hline
local & value & 0.41 & -  &0.45 & - & 0.48& 0.47 & 0.46  \\ \cline{2-9}
min & year & 1981 & -   & 1985 & - & 1967 & 1987 & 1987   \\ \hline
\end{tabular}
\end{center}
\caption{The local extrema  of the mean distance between  the 20 rich countries in the case of the correlation distance for GDP {\it per capita} time series on LMST networks.
\label{tab:gdp_per_capita}}
\end{table}

(ii) {\it Mean Manhattan distance}: for both considered network structures and all considered time windows in the case of the {\it total} GDP time series, a monotonical growth is observed. In brief,  for $T=10, E(d_l \mbox{BMLP})$ and $E(d_l \mbox{LMST})$  go from $ 1.5\cdot 10^6$ to $5.5 \cdot 10^6 $. For $T=15,  E(d_l \mbox{BMLP})$ and $E(d_l \mbox{LMST})$ go from $ 1.5\cdot 10^6$ to $5 \cdot 10^6 $.
$T=15,  E(d_l \mbox{BMLP})$ and $E(d_l \mbox{LMST})$ go from $ 1.6\cdot 10^6$ to $4.6 \cdot 10^6 $. For $T=25,   E(d_l \mbox{BMLP})$ and $E(d_l \mbox{LMST})$ go from $ 1.7\cdot 10^6$ to $4.3 \cdot 10^6 $.

In the case of the GDP {\it per capita} time series the evolution of the mean distance between nodes on the BMLP and LMST  is monotonically growing from $E(d_{l})(1950) \approx 2000$ for  $T=10 $, up to $E(d_{l})(1984) \approx 4200$;  for $T=15 $, $E(d_{l})(1980) \approx 4000$;  for $T=15 $, $E(d_{l})(1975) \approx 3800$; $T=20 $, $E(d_{l})(1970) \approx 3500$;

(iii) {\it Entropy correlation distance}: in the case { \it total} GDP time series: for $(5,10) $,   the mean distances oscillate in [0.4,  0.9], for BMLP, and in  [0.3, 0.7] for LMST. The mean distances for $(10,5)$ oscillate as well but between the values for  BMLP: [0.4,0.9],  and   for LMST in  [0.3, 0.88]. In the case of $(10,10) $ the mean distances between BMLP nodes oscillate in the interval [0.4, 0.78].  Notice that $E(d_{se} \mbox{LMST}) $ has one distinguishable long maximum: it begins in 1959 and lasts till 1963; then  $E(d_{se}) $ is decreasing from 0.7 to 0.4   during two consecutive time windows and remains on this level  till 1987, -- the last point of the evolution. For (15, 15) the mean value evolution begins at $E(d_{se} \mbox{BMLP}) = 0.47$, or  $E(d_{se} \mbox{LMST}) = 0.4$ and is increasing during the next four time windows achieving  a local maximum at $E(d_{se} \mbox{BMLP} ) = 0.85$, or $E(d_{se} \mbox{LMST}) = 0.7$ in 1954. Then the mean distance is decreasing to a minimum  $E(d_{se} \mbox{BMLP}) = 0.26$, $E(d_{se} \mbox{LMST}) = 0.24$  in 1962. In the interval [1962, 1967] the mean distances increase for both networks; for the remaining  time they remain at the level $E(d_{se} \mbox{BMLP}) \approx 0.6$, $E(d_{se} \mbox{LMST}) \approx 0.45$. The extremum points of the time evolution of the mean distance between networks nodes in the case of the time window (15,15) are collected in  Table \ref{tab:ent_cor}.

On the other hand, the analysis of GDP {\it per capita} data allows to make the following observations: for the  BMLP networks the mean distance is very scattered and takes values in the interval:  for (5,10) in $0.55 < E(d_{se} \mbox{BMLP} ) < 0.81$; for (10,5), in $0.46 < E(d_{se} \mbox{BMLP} ) < 0.76$; for (10,10), in  $0.52 < E(d_{se} \mbox{BMLP} ) < 0.74$; for  (15,15), in $0.58 < E(d_{se} \mbox{BMLP} ) < 0.85$.  The mean distance between countries for the LMST network and the time window $15 $ is, as in the case of the BMLP also very scattered taking values in the interval for (5,10), in $0.38 < E(d_{se} \mbox{LMST} ) < 0.65$ and for  (10,5), in $0.23 < E(d_{se} \mbox{LMST} ) < 0.86$. For the two other time window cases one increase, one decrease and a stable evolution of the mean distance between nodes can be distinguished: for (10,10) since 1956 the mean distance is growing till 1961  to $E(d_{se} \mbox{LMST} )(1961)=0.85$ and decreasing to the value $E(d_{se} \mbox{LMST} )(1966)=0.4$; for the (15,15) the first minimum is at $E(d_{se} \mbox{LMST} )(1950)=0.3$ after which the mean distance is monotonically growing to $E(d_{se} \mbox{LMST} )(1955)=0.78$, then decreasing to $E(d_{se} \mbox{LMST} )(1963)=0.35$ and remaining at $\approx 0.4$.

\begin{table}
\begin{center}
% use packages: array
\begin{tabular}{|l|l|c|c|c|c|}
\hline
network &  &  \multicolumn{2}{|c|}{BMLP} & \multicolumn{2}{c|}{ LMST}   \\ \hline
window & sizes &  \multicolumn{2}{|c|}{(15,15) } & \multicolumn{2}{c|}{(15,15)}   \\ \hline
local & value & 0.85  & 0.6  & 0.7  & 0.45  \\ \cline{2-6}
max & year  & 1954 & 1967 & 1957 & 1967  \\ \hline
local & value & 0.47 & 0.26 &0.4 & 0.24   \\ \cline{2-6}
min & year  & 1950 & 1962 & 1950 & 1962   \\ \hline
\end{tabular}
\end{center}
\caption{The local extrema  of the mean distance between    the total GDP time series of countries in the case of the entropy correlation distance used to build   both types of networks for the (15,15) time window combination. 
\label{tab:ent_cor}}
\end{table}

\begin{figure}
 \centering
 \includegraphics[bb=50 50 266 201,scale=0.8]{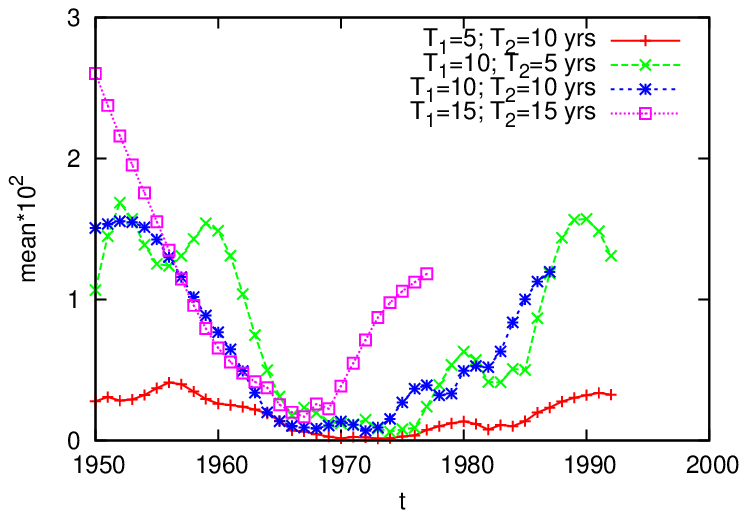}
  \includegraphics[bb=50 50 266 201,scale=0.8]{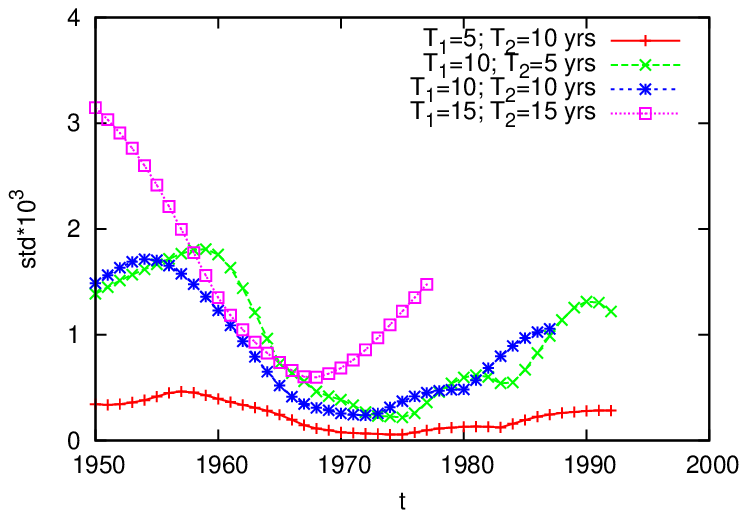}
 % gdp_tot_theil_manh_ewolucja_mean.eps: 0x0 pixel, 300dpi, 0.00x0.00 cm, bb=50 50 266 201
 \caption{(left) The mean distance, (right) the standard deviation between LMST network nodes as a function of time. The LMST networks were constructed from the {\it total} GDP time series through the entropy Manhattan distance concept}
 \label{fig:gdp_mean}
\end{figure}

\begin{figure}
 \centering
 \includegraphics[bb=50 50 266 201,scale=0.8]{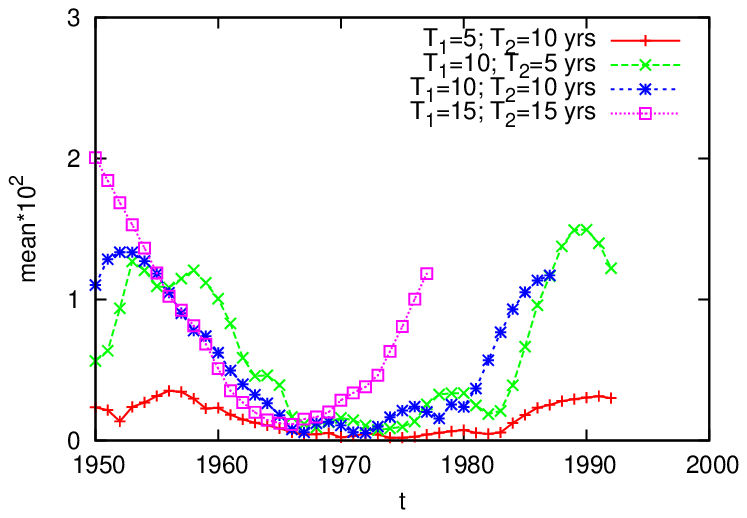}
  \includegraphics[bb=50 50 266 201,scale=0.8]{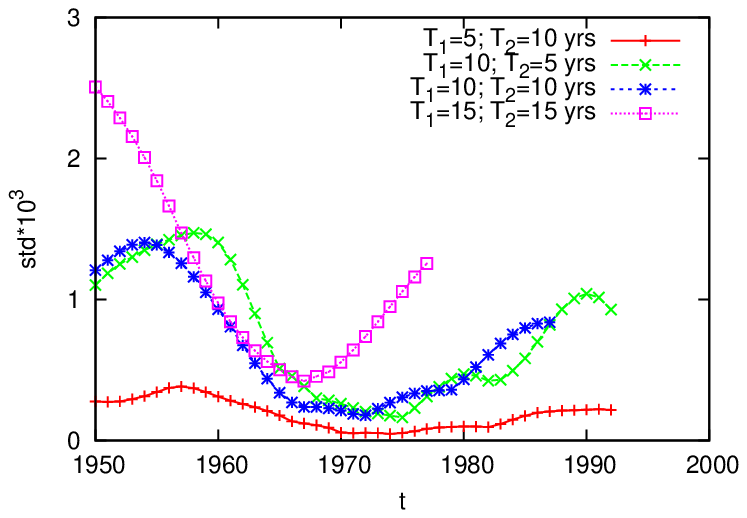}
 % gdp_capita_manh_ewolucja_mean.eps: 0x0 pixel, 300dpi, 0.00x0.00 cm, bb=50 50 266 201
 \caption{(left) The mean distance, (right) the standard deviation between LMST network nodes as a function of time. The LMST networks were calculated from the GDP {\it per capita} time series through the entropy Manhattan distance concept}
 \label{fig:gdp_capita_mean}
\end{figure}

(iv) {\it Entropy Manhattan distance}: as mentioned above we display such results, see Figs. \ref{fig:gdp_mean}-\ref{fig:emp_pop_mean}.    The evolution of the $E(d_{le})$ LMST means are presented in Figs. \ref{fig:gdp_mean}-\ref{fig:gdp_capita_mean} in the case of  the total GDP and the  GDP {\it per capita}, respectively. Since the mean distance between network nodes in the case of the total GDP and the GDP {\it per capita} differ only in a few details, the discussion of its properties of these evolutions concerns both time series together.

 In the evolution of the BMLP and LMST networks three types of evolution can be distinguished: a decrease of the mean distance between network nodes, followed by a period of stable evolution and finally an increase.
In the case of the longest considered time windows (15,15) the decrease period starts in 1950 and ends in 1967 (BMLP and LMST) followed by an increase till 1977; for the time window (10,10) the mean distance is decreasing from 1952 to 1967, then remains stable till 1973 and from that point increases; the evolution of the mean distance between network nodes for the time window (10,5) and (5,10) gives similar results to the discussed cases: a decrease is observed from 1960 till 1967, then the mean distance remains stable until 1976 and from this moment   increases. The main difference between the case (5,10) and (10,5) is that the mean distance is smaller in the time window (5,10), i.e. the maximum of the mean distance is 0.0045 for (5,10)
while it is 0.017  for (10,5). They are 0.0035  and 0.015 for the GDP per capita time series respectively. The BMLP results  concur with the LMST results. 

These observations coincide  with the analysis of the evolution of standard deviation of distances between nodes in LMST networks, as displayed on Fig. \ref{fig:gdp_mean}(b) and \ref{fig:gdp_capita_mean}(b). This indicates that conclusions can logically be reached and make sense when looking at entropy Manhattan distance approach.
 
\subsection{Results/discussion: annually worked hours and employment {\it per capita} ratio}

(i) {\it Correlation distance}: The mean distance between nodes on the BMLP and LMST networks in the case of annualy worked hours is initially growing to a maximum (depending on the time window) in 1955 (for $T=25 $) and 1961 (for $T=10 $) and is later decreasing to a minimum in 1965 (for $T=25 $)) and 1969 (for $T=10 $) before growing for the remanig time of the analysed years.
The analysis of employment {\it per capita} ratio distance matrices  through  BMLP and LMST networks gives very scattered results.  Monotonic evolution periods can be observed only for the LMST network and for the time window $T= 15,$  i.e. in [1972,1975], a decrease and in [1987,1993], an increase.

\begin{figure}
 \centering
 \includegraphics[bb=50 50 266 201,scale=0.8]{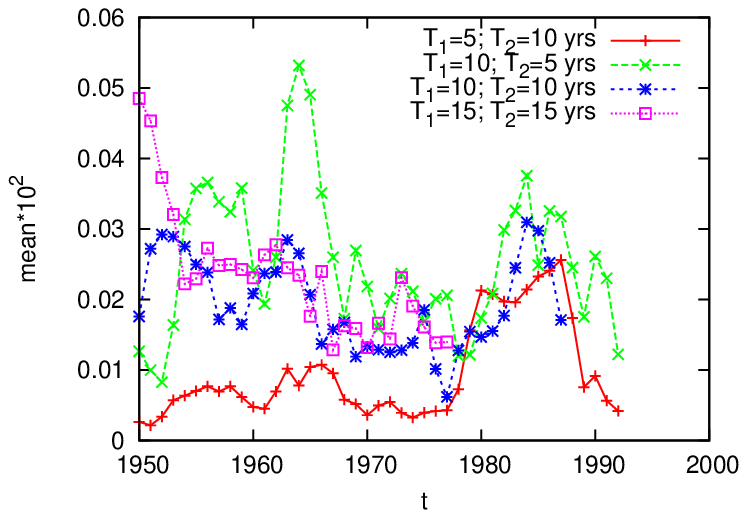}
  \includegraphics[bb=50 50 266 201,scale=0.8]{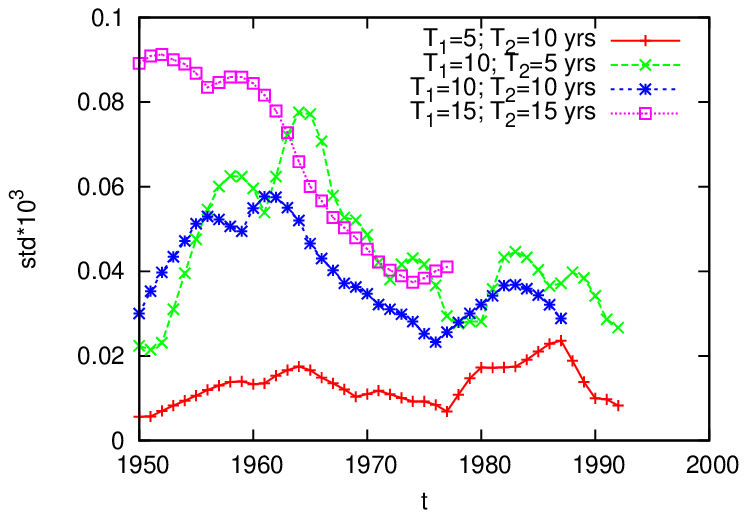}
 % hours_theil_manh_ewolucja_mean.eps: 0x0 pixel, 300dpi, 0.00x0.00 cm, bb=50 50 266 201
 \caption{(left) The mean distance, (right) the standard deviation between LMST network nodes as a function of time. The LMST networks were calculated from the annually worked hours time series through the entropy Manhattan distance concept}
 \label{fig:hours_mean}
\end{figure}

 \begin{figure}
 \centering
 \includegraphics[bb=50 50 266 201,scale=0.8]{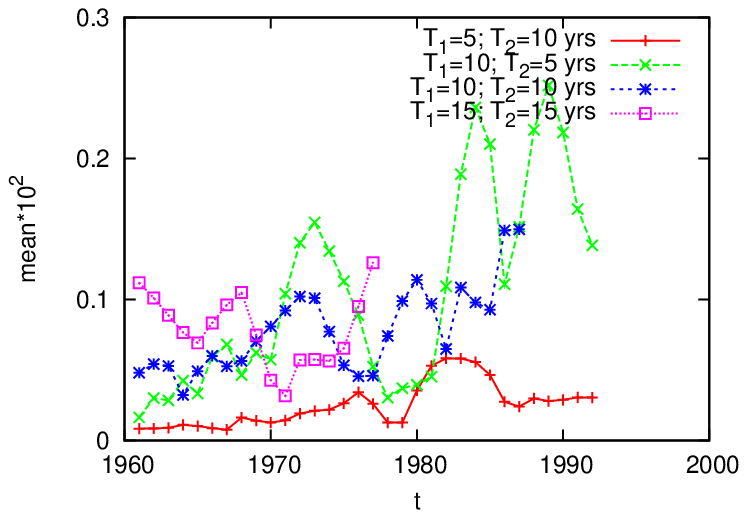}
 \includegraphics[bb=50 50 266 201,scale=0.8]{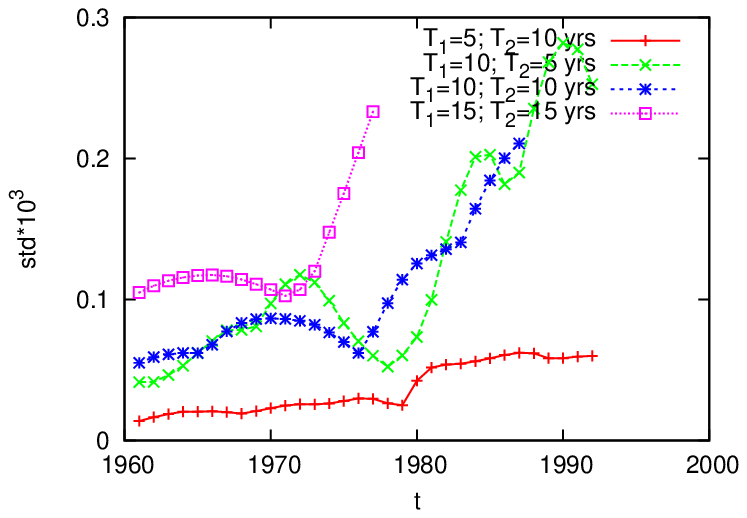}
 % emp_pop_theil_manh_ewolucja_mean.eps: 0x0 pixel, 300dpi, 0.00x0.00 cm, bb=50 50 266 201
 \caption{(left) The mean distance, (right) the standard deviation between LMST network nodes as a function of time. The LMST networks were calculated from the employment {\it per capita} ratio time series through the entropy Manhattan distance concept}
 \label{fig:emp_pop_mean}
\end{figure}

(ii) {\it Mean Manhattan distance}: For the annualy worked hours series, the evolution of the mean distance between nodes on BMLP and LMST networks presents six monotonic periods  with maxima at 1959, 1971, 1984 and minima at 1950, 1965, 1977, 1987 (for $T=10 $). If the time window size is changed, the position of each local extremum is shifted approximately by the increase in the time window size.
For the employment/capita ratio data the mean distances between network nodes increase almost monotonically in the whole considered time interval with one exception, that  of the minimum at 1970  for $T=15 $. Similar evolutions are observed for both networks considered.

(iii) {\it Entropy correlation distance}: An analysis of the annualy worked hours and employment {\it per capita}  ratio results in plots with scattered data points for both network structures in the entropy-correlation distance approach. 

(iv) {\it Entropy Manhattan distance}: The evolution of the mean distance, and its standard deviations,  between LMST network nodes in the case of annually worked hours and employment {\it per capita}  ratio are presented in Figs. \ref{fig:hours_mean}(a)-\ref{fig:emp_pop_mean}(b). Results obtained for BMLP network are very similar to the LMST case.  A monotonic evolution of the mean distances  occurs:  e.g. for the  (5,10) case, a relative stable evolution exists  in the interval [1950, 1977], or for the  (15, 15) case,  a decrease of mean distances occurs in the interval [1950, 1954]. Similar situations are observed in the employment {\it per capita}  ratio data analysis: as in the previous case only a few time intervals can be pointed out as  containing a monotonic evolution of the mean distance: e.g. for the (5,10) a relative stable evolution is found in the interval [1960, 1977].

\section{Conclusions}
\label{sec:concl}

In view of the globalization definition the decrease of the mean distance between countries on  some network is interpreted as the mark of a globalization process.  After constructing appropriate networks, for this macroeconomic analysis,  the evolution of the mean distances, especially their decrease and increase,  convincingly  suggest   globalization and deglobalization periods.

Thus within this paper four distance measures were tested for globalization measures. They were calculated on four sets of time series (i) GDP, (ii) GDP {\it per capita}, (iii) annually worked hours and (iv) employment {\it per capita} ratio. These time series were selected because they are classicaly examined in macro-economy research, and should present a local measure within a globalization process, if any, whence reflecting some integration of the world economy. The second set of time series (iii) annually worked hours and (iv) employment/capita ratio was chosen considering that the globalization process should not {\it a priori} influence this aspect of human activity. In fact the labour market is usually strongly protected by national laws. 

After much calculation, graphs and statistical analysis, we have observed that the most appropriate distance measure is that based on the entropy notion and the best results obtained by calculating the entropy Manhattan distance, based on the Theil index. On one hand,  it shows  a globalization process between considered countries in the case of the total GDP and GDP {\it per capita} time series. On the other hand  the analysis based on the entropy Manhattan distance finds no globalization in the third and fourth data set   (annually worked hours and employment  {\it per capita} ratio). It is thus shown and concluded  that entropy Manhattan distance is interestingly sensitive to measure a globalization process. The calculation of this distance measure allows us to point out periods of globalization {\it and} deglobalization in the world economy. Similar results are obtained for both networks here tested (BMLP and LMST). 
Therefore the choice of the distance measure is seen to be a key factor of  such an analysis, - not the network choice. 

This observation leads to another bonus as a conclusion: since the key factor of the globalization is the increase in the similarities of the entropy evolution, we conclude from the above that the globalization process presents a natural limit. In the case of the 20 countries here above considered this limit was reached in the time interval 1970-2000, as much exemplified by the GDP and GDP  {\it per capita} time series. Of course, we do not pretend that it cannot be reached again, depending on new political conditions, after the presently  apparent deglobalization process.

These numerical and physically based observations concur with European Community considerations about the formation and integration of the considered countries. Especially interesting is the moment of the Berlin wall fall, on  the 9th of November 1989 when new political and economical opportunities arose and a deglobalization process was ``felt'' due to capital flow to the post communist countries. The EUR introduction and the Maastricht agreement constraints seem to indicate the start of the deglobalization. Maybe because they are not followed congruently by the  European countries.

\section*{Acknowledgements}

We would like to thank the organisers of APFA7 and its Satellite workshops for their kind invitation, their warm hospitality and their financial support. MA thanks FRS-FNRS for a travel grant.

\bibliographystyle{elsarticle-num}
\bibliography{apfa_2009_1}

\begin{thebibliography}{10}
\expandafter\ifx\csname url\endcsname\relax
  \def\url#1{\texttt{#1}}\fi
\expandafter\ifx\csname urlprefix\endcsname\relax\def\urlprefix{URL }\fi
\expandafter\ifx\csname href\endcsname\relax
  \def\href#1#2{#2} \def\path#1{#1}\fi

\bibitem{glob2}
R.~Stultz, The limits of financial globalization, The Journal of Finance 60
  (2005) 1595--1638.

\bibitem{glob1}
C.~Russell, The Battle of Armageddon, Bible Students Congregation of New
  Brunswick, 1996.

\bibitem{scholte2000gci}
J.~Scholte, {Globalization: a critical introduction}, Palgrave Macmillan, 2000.

\bibitem{baylis1997gwp}
J.~Baylis, S.~Smith, {The globalization of world politics: an introduction to
  international relations}, Oxford University Press, USA, 1997.

\bibitem{levitt2005gm}
T.~Levitt, {The globalization of markets}, Strategy: Critical Perspectives on
  Business and Management 92 (2005) 399.

\bibitem{hitt2001smc}
M.~Hitt, R.~Ireland, R.~Hoskisson, J.~Parnell, J.~Harrison, C.~John, M.~White,
  G.~Bruton, S.~Abraham, J.~Champoux, et~al., {Strategic management:
  competitiveness and globalization: cases}, South-Western College Pub., 2001.

\bibitem{feenstra1996goa}
R.~Feenstra, G.~Hanson, {Globalization, outsourcing, and wage inequality}, The
  American Economic Review 86 (1996) 240--245.

\bibitem{starr2000nea}
A.~Starr, {Naming the enemy: Anti-corporate movements confront globalization},
  Zed Books, 2000.

\bibitem{dunning1993gb}
J.~Dunning, {The globalization of business}, Routledge London, 1993.

\bibitem{maslov-2001-301}
S.~Maslov, Measures of globalization based on cross-correlations of world
  financial indices, Physica A 301 (2001) 397--406.

\bibitem{mantegna99}
R.~N. {Mantegna}, {Hierarchical structure in financial markets}, Eur. Phys. J.
  B 11 (1999) 193--197.

\bibitem{amin1994gia}
A.~Amin, N.~Thrift, {Globalization, institutions, and regional development in
  Europe}, Oxford University Press, USA, 1994.

\bibitem{chossudovsky2005gp}
M.~M. Suarez-Orozco, D.~B. Qin-Hilliard, {Globalization Culture and Education
  in the New Millenium}, University California Press, Berkely and Los Angeles,
  California, 2004.

\bibitem{parrenas2001sgw}
R.~Parre{\~n}as, {Servants of globalization: women, migration and domestic
  work}, Stanford University Press, 2001.

\bibitem{robertson1992gst}
R.~Robertson, {Globalization: Social theory and global culture}, Sage, 1992.

\bibitem{beyer1994rag}
P.~Beyer, {Religion and globalization}, Sage, 1994.

\bibitem{theil1967eai}
H.~Theil, {Economics and information theory}, Rand McNally, 1967.

\bibitem{mst1}
M.~Eryigit, R.~Eryigit, Network structure of cross-correlations among the world
  market indices, Physica A 388 (2009) 3551--3562.

\bibitem{mst2}
W.-S. Jung, O.~Kwon, F.~Wang, T.~Kaizoji, H.-T. Moon, H.~E. Stanley, {Group
  dynamics of the Japanese market}, Physica A 387 (2008) 537--542.

\bibitem{mst3}
G.~J. Ortega, D.~Matesanz, Cross-country hierarchical structure and currency
  crisis, Int. J. Mod. Phys. C 17 (2006) 333--341.

\bibitem{growth2007dca}
{Development Centre and the Conference Board, Total economy Database},
  http://www.conference-board.org/economics/ (2008).

\bibitem{ppp}
Handbook of The International Comparison Programme., Studies in Methods F, No.
  62, United Nations Publication, 1992.

\end{thebibliography}

\end{document}